\documentstyle[sprocl,epsfig,amss,wrapfig]{article}

\def\be{\begin{equation}}
\def\ee{\end{equation}}
\def\bea{\begin{eqnarray}}
\def\eea{\end{eqnarray}}
\newcommand{\scaption}[1]{\caption{\protect{\footnotesize  #1}}}

\newcommand{\av}[1]{\mbox{$ \langle #1 \rangle $}}

\newcommand{\kjet}{\mbox{$k_{T\rm{jet}}$}}
\newcommand{\xjet}{\mbox{$x_{\rm{jet}}$}}
\newcommand{\ejet}{\mbox{$E_{\rm{jet}}$}}

\newcommand{\ptjet}{\mbox{$p_{T\rm{jet}}$}}

\newcommand{\xb}{\mbox{$x~$}}  

\newcommand{\Qsq}{\mbox{$Q^2~$}}

\newcommand{\et}{\mbox{$E_T~$}}
\newcommand{\kt}{\mbox{$k_T~$}}
\newcommand{\pt}{\mbox{$p_T~$}}

\newcommand{\ftwo}{\mbox{$F_2~$}}

\newcommand{\GeV}{\mbox{\rm ~GeV~}}

\newcommand{\GeVsq}{\mbox{${\rm ~GeV}^2~$}}

\begin{document}

\noindent
{\tt MPI-PhE/97-24 \hfill     September 1997} \\

\title{LOW-$x$ HADRONIC FINAL STATES AT HERA}
\author{ MICHAEL KUHLEN}
\address{Max-Planck-Institut f\"ur Physik, Werner-Heisenberg-Institut\\
 F\"ohringer Ring 6, D-80805 M\"unchen, Germany\\
E-mail: kuhlen@desy.de \\ \hspace{1cm} \\
{\bf
representing the H1 and ZEUS Collaborations at the \\
Madrid Workshop on low-$x$ Physics, Miraflores de la Sierra, June 1997}
}
\maketitle\abstracts{
Measurements of the hadronic final state at HERA are reviewed, which
aim at the investigation of the parton dynamics of the proton
at small Bjorken $x$.
}
\section{Introduction}                      
Amongst the most interesting issues of HERA physics is QCD in the
newly accessible regime of small Bjorken $x$. The observed rise of the
structure function \ftwo towards small $x$ 
suggests a strong
increase of the parton density in the proton,
but what is its dynamical origin?
Is BFKL \cite{th:bfkl} dynamics at work,
or is conventional DGLAP \cite{th:dglap} evolution sufficient?
Complementary measurements of the hadronic final state provide more
detailed information than the inclusive \ftwo data
to help uncover
the underlying dynamics.

\begin{figure}[b]
   \centering
   \vspace{-0.5cm}
   \begin{picture}(0,0) \put(0,0){{\bf a)}} \end{picture}
   \epsfig{file=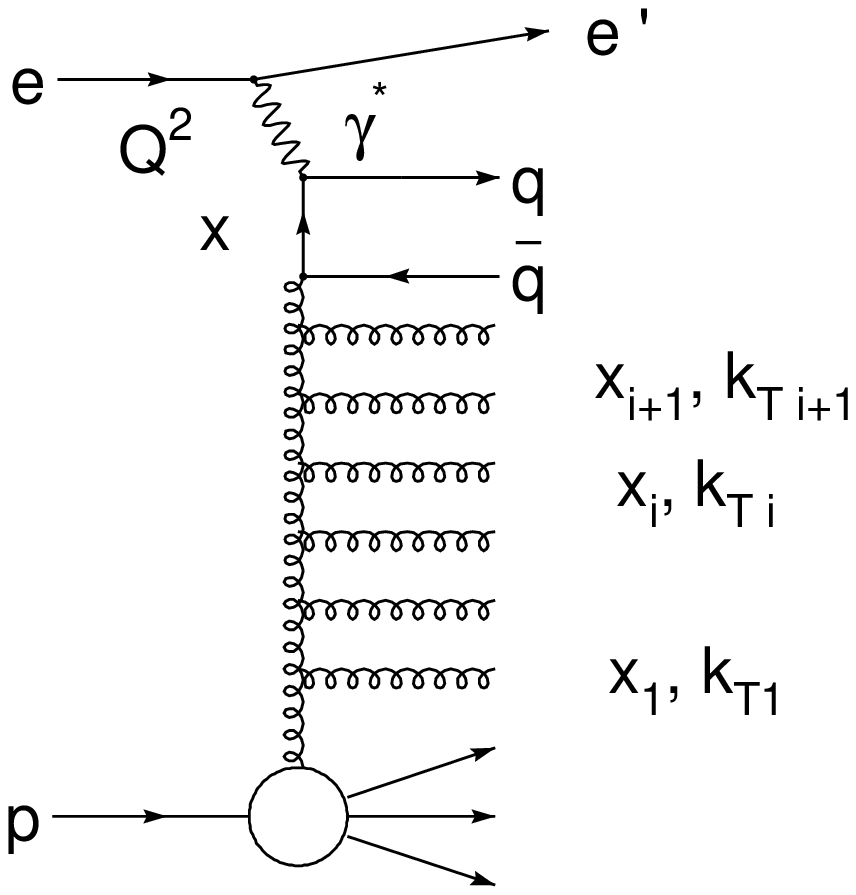,
    width=4cm}
   \hspace{2cm}
   \begin{picture}(0,0) \put(0,0){{\bf b)}} \end{picture}
   \epsfig{file=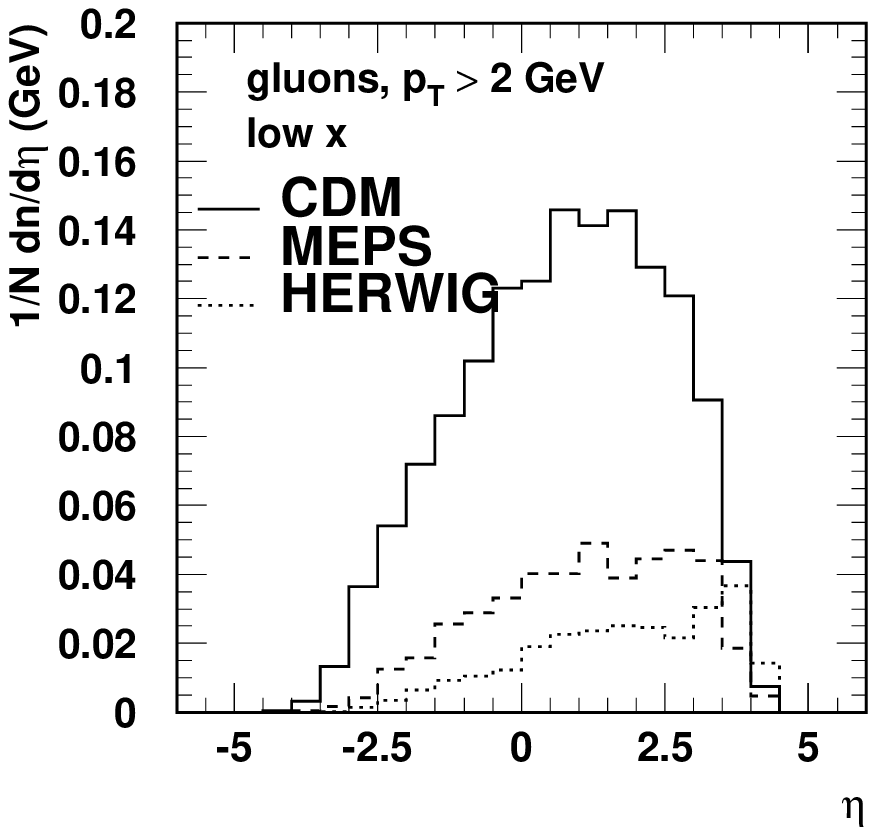,
           width=4cm,bbllx=4pt,bblly=275pt,bburx=257pt,bbury=516pt,clip=}
   \vspace{-0.3cm}
   \scaption{{\bf a)} Ladder diagram for parton evolution.
             {\bf b)}   The multiplicity of hard
      gluons with $\pt>2\GeV$ vs. CMS $\eta$.
      The events are generated with the models
      CDM, MEPS and HERWIG in
      a ``low \xb'' kinematic bin
      with \av{\xb}=0.00037 and \av{\Qsq}=13.1 \GeVsq.
      The proton direction is to the left.}
   \label{ladder}
\end{figure}

The leading log
DGLAP resummation corresponds to a strong
ordering of the transverse momenta \kt (w.r.t. the proton beam)
in the parton cascade,
$Q_0^2 \ll k_{T1}^2 \ll ... k_{Ti}^2 \ll ... Q^2$
(fig.~\ref{ladder}a).
Since in the BFKL evolution that restriction is absent (``\kt diffusion''),
a generic signal for deviations from DGLAP evolution
is enhanced radiation from the ladder between the
current and the remnant system, that is in the central
rapidity region in the hadronic centre of mass system (CMS).
Experimentally pseudorapidity
$\eta = -\ln\tan(\theta/2)$ is used, where $\theta$ is the angle
with respect to the virtual photon axis.
The following observables have so far been exploited and
are being discussed in this paper:
1) \et flows:
increased parton activity should result in an increased
transverse energy flow \cite{lowx:et}.
2) Charged particle \pt spectra:
high \kt partons, disfavoured by the strong \kt ordering
in DGLAP, are signalled by measureable high \pt hadrons
\cite{mk:special}.
3) Forward jets:
high energy jets with $\ptjet^2 \approx \Qsq$
(kinematically
bound to be measured in the forward calorimetric systems
close to the remnant direction)
tag events with BFKL evolution, because DGLAP
evolution is not allowed \cite{lowx:fwdjets,lowx:hotref}.

Theoretical calculations for these observables exist
and can be compared to the data, provided that hadronization effects
can be controlled.
Alternatively,  predictions
are derived from Monte Carlo models, which
incorporate the QCD evolution in
different approximations and utilize phenomenological models
for the non-perturbative hadronization phase.
The MEPS model (Matrix Element plus Parton Shower,
program LEPTO \cite{mc:lepto}) and the HERWIG generator \cite{mc:herwig}
are based upon leading log DGLAP parton showers, with strong
\kt ordering of the emitted partons.
In the colour dipole model (CDM) \cite{mc:dipole,mc:ariadne}
gluon emission is not subject to \kt ordering \cite{mc:bfklcdm}.
In that respect it mimicks the BFKL evolution, and leads to
more abundant gluon radiation than in the other models
(fig.~\ref{ladder}b).

\section{Energy flows}                            

\begin{figure}[tb]
   \centering
   \epsfig{file=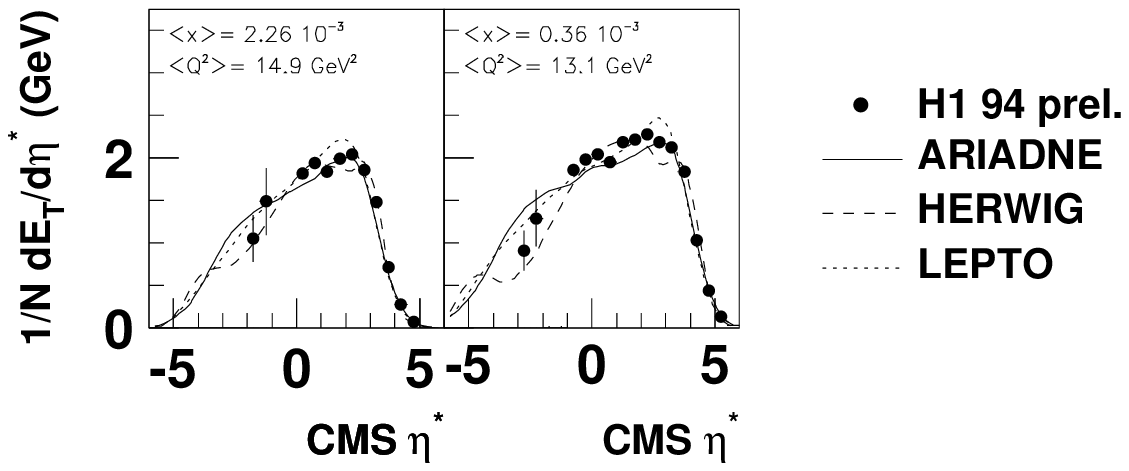,
           width=10cm,bbllx=44pt,bblly=527pt,bburx=370pt,bbury=672,clip=}
   \vspace{-0.4cm}
   \scaption{The \et  flow vs.
             $\eta$ in the hadronic CMS for $\av{x}\approx 0.002$
             and $\av{x}\approx 0.0004$ with $\av{Q^2}\approx 14\GeVsq$
             fixed.
             The proton
             direction is to the left.
             The data \cite{h1:flow4}
             are compared to
             the models
             CDM (ARIADNE 4.08), MEPS (LEPTO 6.4), and HERWIG 5.8.}
   \label{etflow}
\end{figure}

The flow of transverse energy $E_T$
as a function of $\eta$
provides a very simple,
global characterization of the hadronic final state.
Though their partonic radiation patterns are very different,
all models in their present incarnations provide a reasonable
description of the detailed \et flow data by H1 \cite{h1:flow4}
(two out of 17 $x-Q^2$ bins are shown in fig.~\ref{etflow}).

\begin{figure}[tb]
   \centering
   \epsfig{file=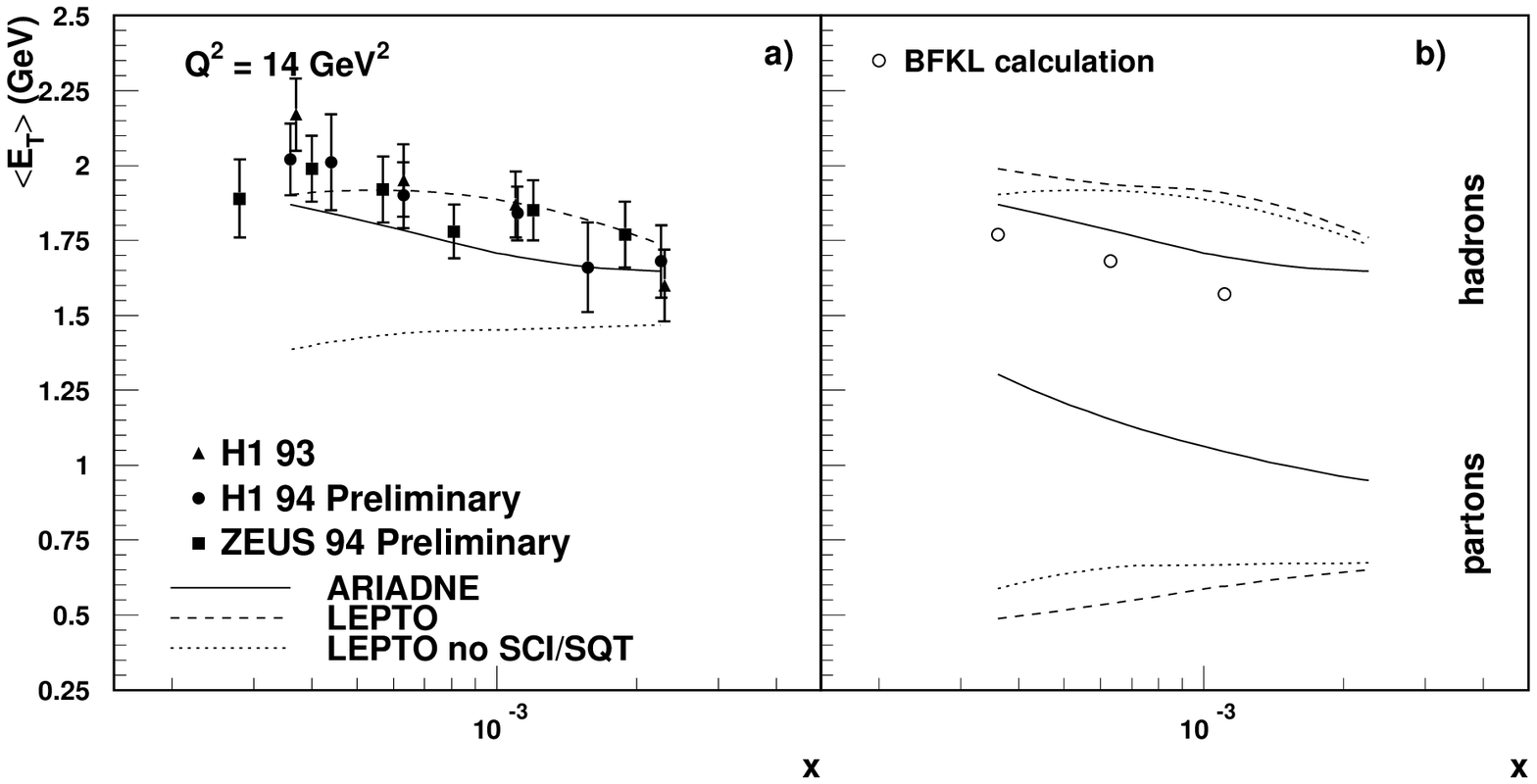,width=12cm}
   \vspace{-0.4cm}
   \scaption{
             The mean \et (GeV) in $-0.5<\eta<0.5$
             as a function of \xb for $\Qsq=14\GeVsq$.
             The data \cite{h1:flow4,z:etdis96} are shown together with
             the models CDM (ARIADNE 4.08), MEPS (LEPTO 6.4), and HERWIG 5.8
             for hadrons and for partons, and with
             the BFKL calculation for partons \cite{lowx:et}.
             The LEPTO result for hadrons without the features
             soft colour interaction (SCI) and the new sea quark
             treatment (SQT) is also shown.}
   \label{etfix}
\end{figure}

From BFKL evolution a relatively large amount of \et
at central rapidity is expected
that increases with decreasing $x$, opposite to the DGLAP expectation
\cite{lowx:et}. These behaviours are realized at the parton level
for the models with the unordered and ordered emission scenarios
(fig.~\ref{etfix}).
The average \et ($\av{E_T}$) measured in
$-0.5<\eta<0.5$ \cite{h1:flow4,z:etdis96}
does increase with falling $x$, in qualitative
agreement with the BFKL expectation (fig.~\ref{etfix}).
However, when comparing to the calculation
hadronization effects have to be taken
into account.

The DGLAP based models have to employ large hadronization corrections
to achieve the level of \et seen in the data (see fig.~\ref{etfix}b).
For example,
in LEPTO the new concept of soft colour interactions (SCI)
has been introduced, resulting in a reconfiguration of
the fragmenting strings, that may lead to an enhancement of \et
(and also to rapidity gaps) \cite{mc:sci}.
In the CDM, hadronization
effects are much smaller than in LEPTO and HERWIG (fig.~\ref{etfix}).
In summary, though the data follow the trend expected from BFKL
evolution and are consistent with a model with an unsuppressed
radiation scenario, they
are also consistent with DGLAP
evolution, assuming large hadronization effects.

\section{Charged particle transverse momenta}    
Not yet well understood hadronization effects precluded strong
conclusions on the underlying parton dynamics from
the \et flow measurements.
Single particle
transverse momentum (\pt) spectra
represent a more direct measure of the partonic
activity \cite{mk:special}.
Unsuppressed
parton radiation should manifest itself in a hard tail of the \pt
distribution,
whereas hadronization should produce typical spectra
limited in \pt.

\begin{figure}[thb]
   \centering
\begin{picture}(0,0) \put(0,0){{\bf a)}} \end{picture}
   \epsfig{file=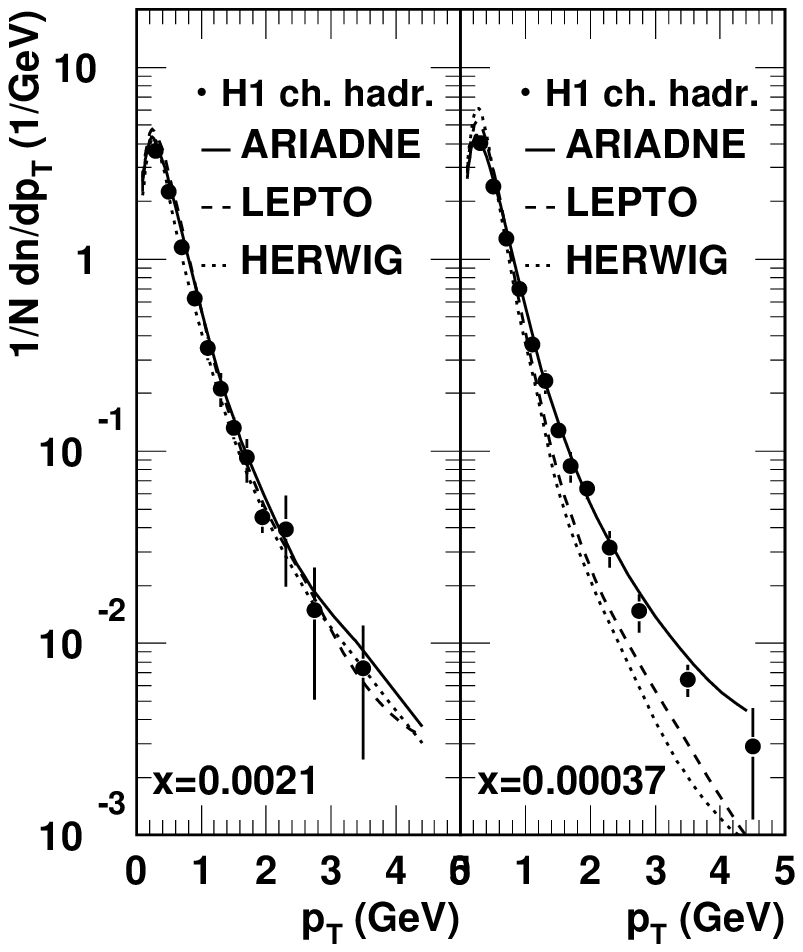,%
    width=5.9cm}
\hspace{-0.5cm}
\begin{picture}(0,0) \put(0,0){{\bf b)}} \end{picture}
   \epsfig{file=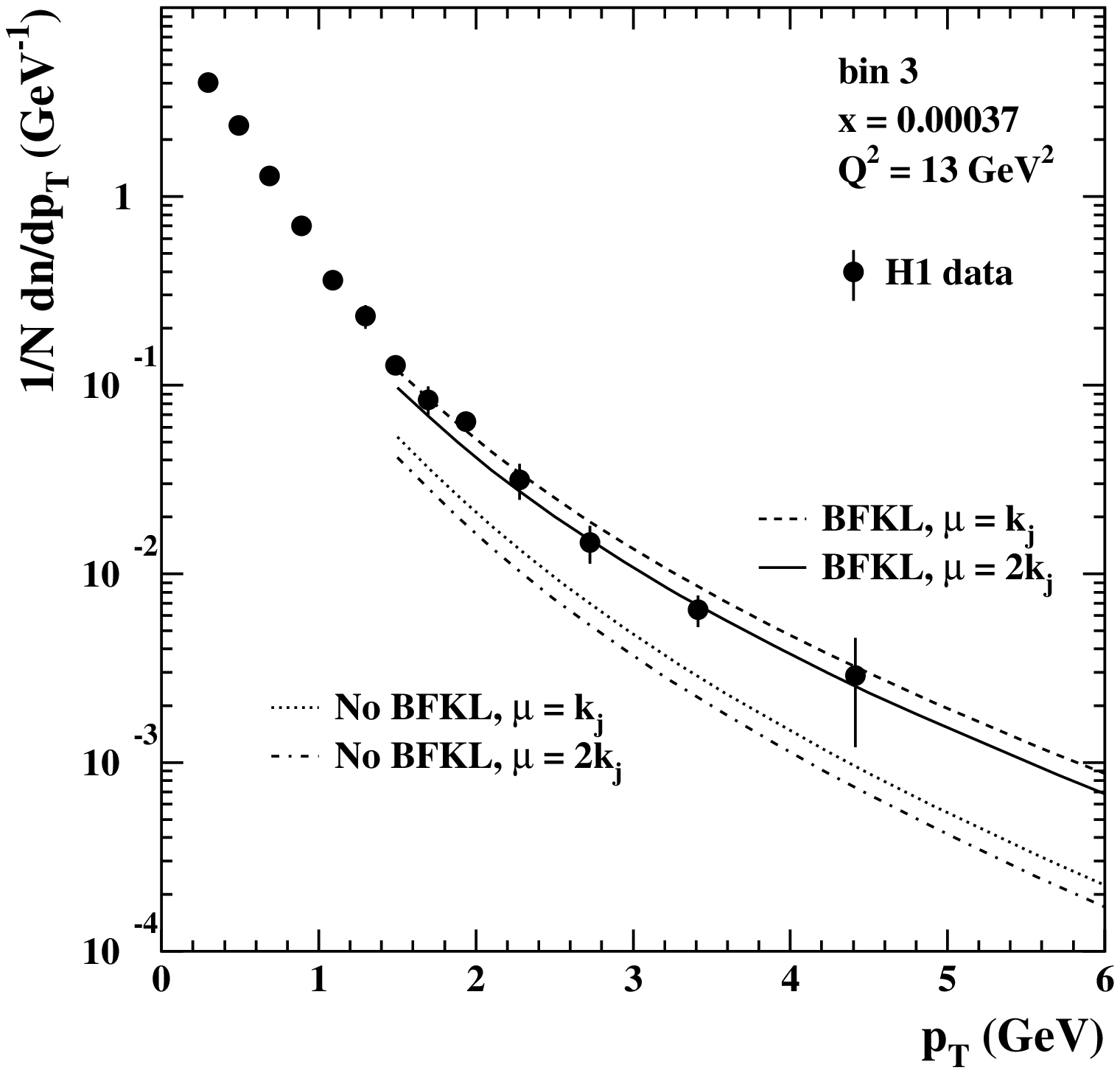,
    width=5.9cm,bbllx=49pt,bblly=210pt,bburx=490pt,bbury=645pt,clip=}
   \vspace{-0.4cm}
   \scaption{
            {\bf a)} The \pt spectra of charged particles from
               $0.5 < \eta < 1.5$ (CMS) \cite{h1:pt}.
               Displayed are two different kinematic bins
               at high and low $x$ for $\av{Q^2} \approx 14\GeVsq$.
               The models
               ARIADNE 4.8, LEPTO 6.4 and
               HERWIG 5.8 are overlayed.
             {\bf b)} The result of a theoretical calculation with and without
                BFKL evolution \cite{lowx:pt},
                compared to the H1 data at low $x$
                for different choices of the factorization scale
                $\mu$.}
   \label{ptcalc}
\end{figure}

H1 has measured
the charged particle \pt spectra as much central
in the CMS
($0.5<\eta<1.5$)
as the tracking detector acceptance allowed \cite{h1:pt}.
At large $x$, all models agree with the data, but at small
$x$, only the model with the unsuppressed radiation pattern (CDM)
is able to describe the
high \pt tail seen in the data (fig.\ref{ptcalc}a).
The shortfall
of the models with suppressed gluon radiation indicates that
at small \xb there is more high $k_T$ parton radiation present than is
produced by the models based upon leading log DGLAP parton showers.

The \pt spectra at small $x$ have been calculated by
folding a cross section $\sigma_j$ to produce a parton $j$
with experimentally known fragmentation functions $D_{h/j}(z)$
to produce a hadron $h$ with momentum fraction $z$ from the parton $j$
\cite{lowx:pt}.
Symbolically,
   $\sigma_h = \sigma_j \otimes D_{h/j}$.
Monte Carlo assumptions for hadronization are thus avoided.
When BFKl evolution is invoked, the H1 data are well described
by the calculation (fig~\ref{ptcalc}b).
The normalization for the BFKL part
is obtained by requiring that the calculated parton cross
section would match the measured H1 forward hadron jets
\cite{h1:fwdjet} (see next sect.).
Neglecting BFKL evolution, the calculation
falls significantly below the data.
It would now be interesting
to compare a complete
fixed order calculation in next-to-leading order (NLO)
to the data.

\section{Forward jets}                          
Forward jets are
the classic signature for BFKL evolution
\cite{lowx:fwdjets,lowx:hotref}.
One requires $\xjet=\ejet / E_p$, the ratio of jet and proton energy
to be large in order to maximize the BFKL evolution from \xjet~ down to $x$.
The jet transverse momentum \kjet has to be close to $Q$,
$\kjet \approx Q$
to suppress the phase space for DGLAP evolution.
In the presence of BFKL evolution,
the forward jet cross section should grow faster with decreasing $x$
than for DGLAP evolution.

\begin{figure}[tb]
   \centering
   \epsfig{file=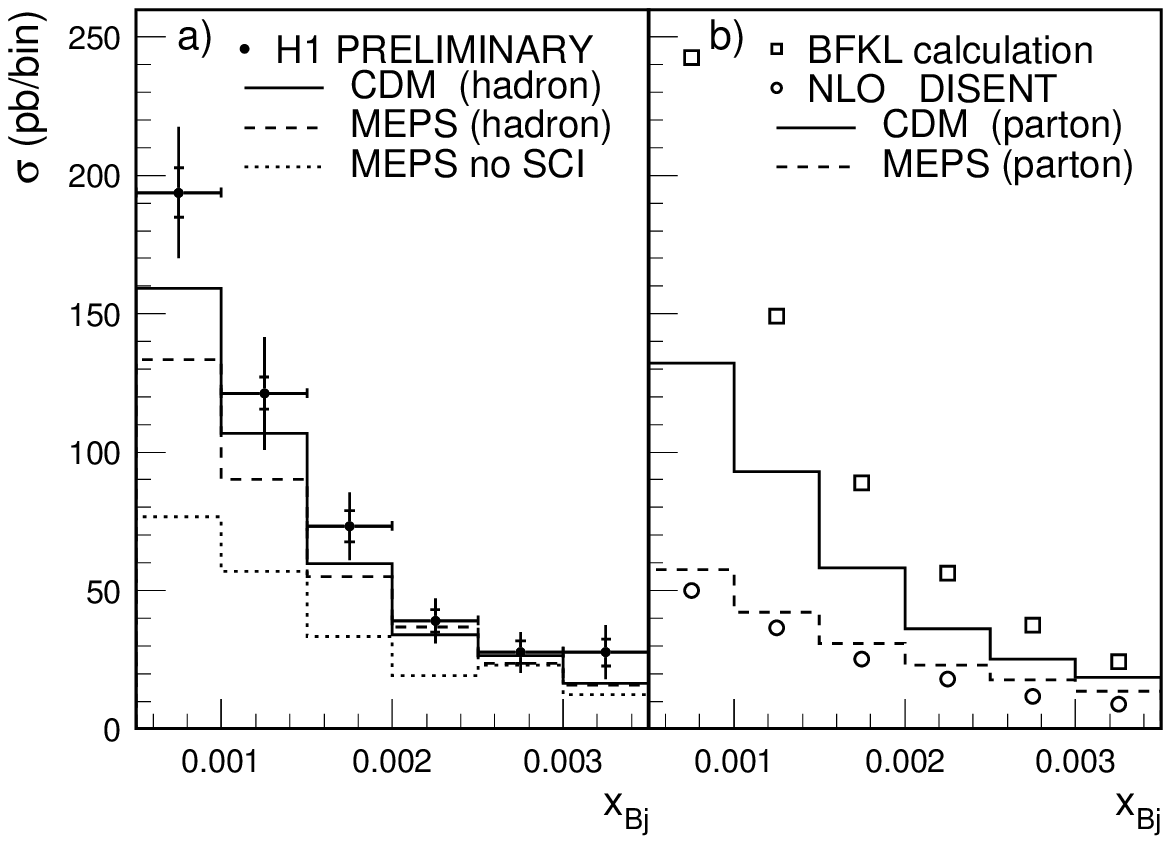,
           width=7.6cm}
   \hspace{-0.2cm}
   \begin{picture}(0,0) \put(0,0){{\bf c)}} \end{picture}
   \epsfig{file=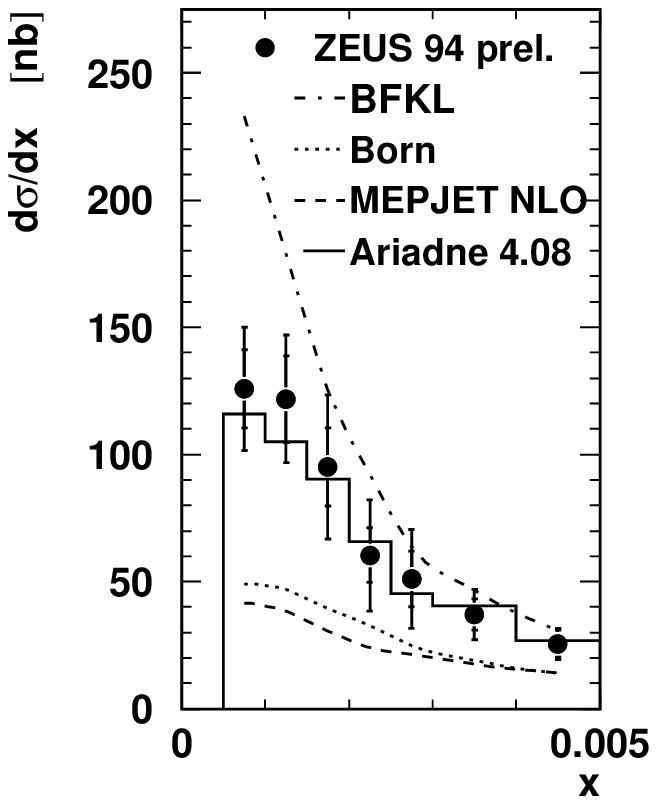,%
           width=4.2cm,bbllx=37pt,bblly=416pt,bburx=226pt,bbury=645,clip=}
   \vspace{-0.4cm}
   \scaption{{\bf a)}
             The H1 forward jet cross section \cite{h1:fwdjet} (1994 data)
             vs. \xb, corrected for detector effects to the hadron level.
             $\xjet>0.035$, $0.5 < \ptjet^2/Q^2 < 2$ and $\ptjet >$ 3.5 \GeV
             were required.
             Also shown are the hadron level predictions
             from CDM (ARIADNE 4.08) and MEPS
             (LEPTO 6.4, with and without SCI).
             {\bf b)}
             The corresponding parton level forward jet cross section
             from a NLO
             calculation \cite{mc:disent}, and for CDM and MEPS.
             Also shown is a BFKL parton cross section calculation
             \cite{lowx:fwdcalc2} (no jet algorithm).
             {\bf c)}
             The ZEUS forward jet cross section \cite{z:fwdjet} (1994 data)
             vs. \xb, corrected to the parton level.
             $\xjet>0.035$, $0.5 < \ptjet^2/Q^2 < 4$ and $\ptjet >$ 5 \GeV
             were required. The data are compared to a NLO jet calculation
             \cite{mc:mepjetz} and to parton jets from CDM (ARIADNE 4.08).
             Also shown are parton cross sections (no jet algorithm)
             with (``BFKL'') and without (``Born'') BFKL evolution
              \cite{lowx:fwdcalc2}. The systematic errors do not
             include uncertainties due to hadronization.}
   \label{fjet}
\end{figure}

H1 reconstructs forward jets with the cone algorithm and requires
$\ptjet >$ 3.5 \GeV \cite{h1:fwdjet}.
The forward jet cross section (fig.~\ref{fjet}a),
corrected for detector effects,
is well described by the CDM, and somewhat less well by the
standard MEPS model with SCI (note that SCI was necessary
in order to describe the \et flows).
Without SCI, the DGLAP based model MEPS cannot
describe the growth of the cross section towards small $x$.
Comparing the hadron level jets to the
parton level jets (fig.~\ref{fjet}b), one finds
relatively small hadronization corrections in the CDM and
also in MEPS without SCI, but with SCI,
hadronization effects become large!
If one would assume the CDM
hadronization corrections to be correct,
the data would be far above both the NLO calculation \cite{mc:disent}
and the parton level jets from the DGLAP based model MEPS.

Presumably hadronization uncertainties become smaller
for a higher \ptjet \\ \mbox{cut-off}.
In the ZEUS analysis \cite{z:fwdjet},
which relies also on the cone algorithm,
$\ptjet >$ 5 \GeV is required.
The data in fig.~\ref{fjet}c
are corrected to the parton level, assuming hadronization
corrections from the CDM.
In contrast to a NLO jet calculation \cite{mc:mepjetz},
the CDM describes the data well.

It is clear that in order to draw firm conclusions from
the forward jet analyses,
hadronization effects need to be better understood.
It is however interesting to note that the
BFKL calculation \cite{lowx:fwdcalc2} is far above
the Born level calculation (excluding BFKL evolution), and
describes qualitatively
the rise of the measured forward jet cross sections at small $x$
(fig.~\ref{fjet}).
Direct comparison to the data however would be imprudent,
because these calculations do not invoke a jet algorithm,
and hadronization corrections are potentially large.




\section{Conclusion}                         
The transverse energy flow and the forward jet data
are only compatible with a conventional DGLAP
evolution scenario when large hadronization effects
are assumed.
The high \pt tail seen in the charged particle spectra
cannot be explained with hadronization effects.
A theoretical calculation neglecting BFKL evolution,
where hadronization is taken into account via known
fragmentation functions, falls far below the data.
The \pt spectra require more hard parton radiation than
expected from conventional DGLAP evolution.
BFKL effects offer a consistent explanation of the
measured \pt spectra, forward jets and \et flows.
Of course this does not
exclude other explanations.
For example, it has been
suggested that contributions from resolved virtual photons
could be responsible for enhanced
parton activity \cite{lowx:madridjungm}.
In that case
the hard scattering takes place between partons
inside the resolved photon and partons from the proton
and may thus happen not at the photon vertex but
further down the ladder.

\section*{References}
\begin{footnotesize}
%
%
%
%
%

\end{footnotesize}
\end{document}